\documentclass[aip,pop]{revtex4}
\newcommand{\beq}{\begin{equation}}
\newcommand{\eeq}{\end{equation}}
\usepackage{graphics}
\usepackage{epstopdf}
\usepackage{epsfig}

\begin{document}

\title{Simulations Reveal Fast Mode Shocks in Magnetic Reconnection Outflows} 
\author{Jared C. Workman}
\affiliation{Department of Physics \& Astronomy, University of Rochester, Rochester NY, 14627}
\affiliation{Department of Physical \& Environmental Sciences, Colorado Mesa University, Grand Junction CO, 81501}
\author{Eric G. Blackman}
\affiliation{Department of Physics \& Astronomy, University of Rochester, Rochester NY, 14627}
\affiliation{Laboratory for Laser Energetics, University of Rochester, Rochester NY, 14623 }
\author{Chuang Ren}
\affiliation{Department of Physics \& Astronomy, University of Rochester, Rochester NY, 14627}
\affiliation{Laboratory for Laser Energetics, University of Rochester, Rochester NY, 14623 }
\affiliation{Department of Mechanical Engineering, University of Rochester, Rochester NY, 14627} 

\begin{abstract}
\
Magnetic reconnection is commonly perceived to drive flow and particle acceleration in   flares of
 solar, stellar, and astrophysical disk coronae but the relative roles of different acceleration mechanisms  in  a given reconnection environment are not well understood.   While 
   outflow fast mode shocks have been predicted analytically, 
   we show for the first time via direct numerical simulations that such  shocks do  indeed occur
   in the outflows of fast reconnection when an obstacle is present. These shocks are 
 are distinct from the slow mode Petschek inflow shocks.   If Fermi acceleration of electrons operates in the weak fast shocks the associated compression ratios will induce a Fermi acceleration particle spectrum that is significantly steeper than  strong fast shocks commonly studied, but  consistent with the demands of solar flares. While this is not the only acceleration mechanism operating in a reconnection environment, it is plausibly a ubiquitous one.

\end{abstract}

\maketitle
\date{\today}

\section{Introduction}

Magnetic reconnection is a frequently studied  process in plasma  astrophysics whereby
magnetic energy is converted into some combination of thermal, non-thermal, and flow kinetic energy.  Reconnection  underlies  models of solar and stellar flares and coronae of astrophysical accretion disks. The process  has also been studied in laboratory plasma experiments \cite{yamadakulsrud}.

Most  work  has focused on  understanding the reconnection rate  \cite{birn01}  
and under what circumstances fast reconnection occurs.  Several distinct paradigms for the onset of
fast reconnection in nature have been studied and may operate in complementary environments.
The first occurs  when the ion inertial length is larger than the magnetohydrodynamic (MHD) Sweet-Parker resistive length,  facilitating current driven plasma instabilities that enhance  dissipation near the X-point \cite{biskamp00,kulsrud01,uzdensky03}.
A second  occurs in the context of  turbulent MHD, where the global  reconnection rate is enhanced by the contemporaneous action of many small scale sites \cite{loureiro07,kowal09,eyink11}.
A third path to fast reconnection occurs for a long enough laminar MHD current sheet that becomes tearing mode unstable
and  again multiple small scale reconnection sites acting together enhance the rate.
\cite{bhy2009}.

There has been  less  synthesized progress understanding how the converted magnetic energy
partitions and on the  accelerated particle  spectra although different mechanisms have been investigated\cite{romanovalovelace92,blackman1,blackman96,dalpino,kowal,blackman97,tsuneta97}.
Because of the variety of  plasma  reconnection conditions, it is best to think of reconnection as an acceleration environment rather than a single  mechanism. Different processes can operate with different strengths depending on the conditions, and whether reconnection is fast or slow.

The basic question we address here, using numerical simulations, is whether  fast shocks arise generically in reconnection outflows. Analytic predictions and phenomenological scenarios suggest that weakly compressive fast shocks may form downstream of reconnection sites  \cite{blackman1,blackman97, tsuneta97,tsuneta98}, such as  in coronal flares if  supersonic  flows from the X-point impinge on a previously reconnected loop top from above.  However, there has  yet  to be a study of this specific such shock formation numerically. Note that the fast
outflow shocks we are focusing on are distinct from slow mode inflow shocks of the Petschek model \cite{petschek}.
Even weak fast shocks  in reconnection outflows would be important because of their potential ubiquity and  role for shock-Fermi acceleration \cite{bell2,JandE}.    A fast downstream shock in solar flares \cite{blackman1,tsuneta98,blackman2}  also benefits from magnetic dissipation near the X-point that may  provide  the needed injection electrons that seed the Fermi process.

Here we present 2-D magnetohydrodynamic (MHD) simulations of initial Harris current sheet configurations  that are unstable to magnetic reconnection, and determine the conditions for which  the outflows are weakly super fast-magnetosonic. By injecting a dense plasma obstacle into the downstream, we then directly show the formation of  weak outflow shocks.  From the numerically computed compression ratios, we discuss the implications for  electron energy spectra from shock  Fermi acceleration and discuss how the results may apply to solar flares.
 
\section{Methods}

For our simulation we use the MHD code ATHENA  \cite{athenamethod,athenamethod2}, 
a cartesian, time explicit, unsplit, Godunov, code parallelized by MPI for compressible MHD.     The mass conservation, momentum,  and magnetic induction equation to be used are given by 
\begin{eqnarray}
\frac{\partial \rho}{\partial t} +
{\bf\nabla\cdot} \left(\rho{\bf v}\right) & = & 0
\label{consmass} \\
\frac{\partial (\rho {\bf v})}{\partial t} +
{\bf\nabla\cdot} \left(\rho{\bf vv} - {\bf BB}\right) +
{\bf \nabla} P^{*} & = 0
\label{consmom} \\
\frac{\partial {\bf B}}{\partial t} +
{\bf\nabla}\times \left({\bf v} \times {\bf B}+\nu_m{\bf J}\right) & = & 0 
\label{bfield},\\
\nonumber
\end{eqnarray}
where 
$P^* \equiv P + ({\bf B \cdot B})/2$,
$P$ is fluid pressure, $\mathbf{B}$ is the magnetic field, $\rho$ is the mass density,  $\mathbf{v}$ is the velocity field,  and the magnetic diffusivity $\nu_m$ equals the resistivity, i.e.   $\nu_m={\eta_m}$. In the units used in ATHENA the magnetic permeability is  unity and one converts to cgs units by dividing $\mathbf{B}$  by $\sqrt{4 \pi}$.

In Eq. (\ref{bfield}) ${\bf J} = {\bf\nabla} \times {\bf B}$, and we write $\nu_m$ as
\beq
\nu_{m}=(\nu_{c}+\nu_{e} e^{\frac{r^2}{2\sigma^2}}),
\label{eta}
\eeq
\noindent where $\nu_c$ is a uniform background resistivity and 
 $\nu_e$ is an enhanced resistivity. 
 The quantity $\sigma$  determines the radial scale  of the enhanced resistivity about the simulation origin  surrounding the X-point.  The functional form of the gaussian fall off of this enhancement
 is arbitrary and is simply chosen to provide the option to consider the effect of  enhanced dissipation near the X-point for $\nu_e>0$, the importance of which we will discuss later.

We set up initial isothermal equilibria for all cases but then distinguish cases in which 
perturbations are evolved adiabatically or isothermally. These cases are referred to as "adiabatic"
and "isothermal" respectively. Note that by adiabatic we mean no heat loss from the system. Neither the adiabatic or isothermal cases are isentropic.
As the isothermal and adiabatic simulations  bracket a wide range of cooling efficiencies for the perturbed configurations  motivates our inclusion of both cases.

For both adiabatic and isothermal simulations we  initialize $P$ as 
\begin{equation}
P =\rho c_s^2,
\label{press}
\end{equation}
where $c_s$ is the sound speed and is set equal to .5 initially and $\rho$ is initialized as discussed later in this section.

For our isothermal simulations no energy equation is solved and Eq. (\ref{press}) becomes a fixed definition.  For our adiabatic simulations  we replace Eq. (\ref{press}) with the equation for   energy evolution given by 
\begin{equation}
\frac{\partial {\bf U}}{\partial t}= {\bf\nabla\cdot} [(\bf U +P^{*})\bf v - \bf B (\bf B \cdot \bf v)+\nu_m \mathbf{J} \times \mathbf{B}] = 0,
\label{energy}
\end{equation}
where
\beq 
\bf U = P/(\gamma - 1) + (\rho \bf v \cdot \bf v /2)+(\bf B \cdot B)/2.
\eeq
  and $c_s$  evolves as $\sqrt{\gamma P/ \rho}$ with $\gamma =5/3$ and evolves as $P$ evolves according to Eq. (\ref{energy}). 

Our initial conditions follow  those in the subset of the MHD  GEM challenge simulations\cite{birn01} with only resistivity and no viscosity. Technically, in MHD the gyro-radii and plasma particle inertial lengths are  infinitesimal, and the  non-ideal nature of the plasma is scalable entirely with 
 the Lundquist or Magnetic Reynolds numbers given by $L v_a/\nu_M$ where $v_a=B \sqrt{4\pi\rho}$ is the Alfv\'en speed. 
 
It is tempting to  normalize lengths  to the ion inertial length $c/\omega_{pi}= c/\sqrt{4\pi n_0e^2/m_i}$ and our time scale to the inverse ion gyrofrequency $(eB/m_ic)^{-1}$ to see what kind of scales   our choices of number density and magnetic field, and temperature would imply 
for a real system as well as to compare to the MHD and particle in cell simulations of the GEM Challenge\cite{birn01}.  
 In our simulations, $\hat {\bf x}$ 
 is the inflow axis and $\hat {\bf y}$  is the axis of  the current sheet.  We employ a rectangular domain with $-L_x/2 \le x \le L_x/2$ and $-L_y/2 \le y \le L_y/2$, where $ L_x = 12.8$ and $ L_y = 25.6$
 in arbitrary units.    Upstream conditions of  a particular solar flare \cite{tsuneta},
 $n_0 = 10^9 {\rm cm^{-3}}$,  and $B_0=20 $G, 
   corresponds to an  ion inertial length of $720$cm, ion gyroradius of $29$cm, and  Alfv\'en speed  $v_A\sim 1378{\rm km\ s^{-1}
 }$ so this indicates that that for this flare we (and the authors of  other simulations\cite{birn01})  would be assuming MHD is valid for scales $> 12.8\times 720$cm.  This is misleading however, since the presence or absence of shocks once in the MHD regime does not depend on the actual normalization of the density or the magnetic field. One need only be concerned that the MHD approximation is valid
 on scales for which a comparison to a real system is being made and to check the
 correpsonding Lundquist number.

Resolutions for the runs in Table 1 range from $384 \times 768$ to $960 \times 1920$.  Most simulations were run at $480 \times 960$.  The primary effect of resolution was to change the steepness of the flow profiles near the x point but not the final magnitude of the outflow at the simulation edges.
Outflow boundary conditions were used at all edges for both  fluid and magnetic quantities. 
The  mass and magnetic field  were not resupplied as the simulations evolved.
Perturbation around a stable configuration  causes the commencement of reconnection, which   draws material  toward the y-axis.

ATHENA maintains a zero divergence of the magnetic field if the initial divergence is zero. To ensure $\nabla\cdot {\bf B}=0$ in ATHENA, we 
 initialize the magnetic field via  a vector potential $\bf A$ such that $\nabla\times {\bf A}={\bf B}$.
For our unperturbed Harris configuration we used  
\beq
{A_z}(x) = -\frac{b_0}{a}\log{(\cosh{(x/a)})\mathbf{\hat z}},
\eeq
where $b_0 = 1.0$ and $a=0.5$ for all simulations which then returned 
\beq
{B_y}(x)= -\frac{\partial {A_z}}{\partial x}=b_0\tanh{(x/a)}.
\label{mainfield}
\eeq
The density profile for our initial Harris configuration is
\beq
\rho(x)=\frac{\rho_0}{\cosh^2{(x/a)}}+\rho_\infty,
\eeq
where $\rho_0 = 1$ and $\rho_\infty=0.2$ for all simulations.  Inclusion of a constant background density was used to prevent a large an Alfv\'en speed in low density regions that  would result in an unreasonably small time step.  The overall pressure balance and dynamics were left unchanged by the background density.

Using $c_s = \sqrt{0.5}$,  makes $P^*={b_0^2}/{2}+{\rho_\infty c_s^2}$, and $\beta={b_0^2/(2\rho_\infty c_s^2)}= 0.2$  and represents an initial equilibrium  around which we peturb. 
For the cases we call "adiabatic"  we set the initial
  unperturbed  $P$ in Eq. (\ref{energy}) to the isothermal equilibrium pressure, but then set $\gamma = 5/3$.
The deviation from equilibrium then evolves adiabatically according to Eq. (\ref{energy}).

To initiate reconnection, we use a vector potential perturbation 
\beq
{A_z'}(x,y)= -\psi_0\cos(kx)\cos(ky)
\label{perturb}
\eeq
 where $k=\frac{\pi}{L_x}=\frac{2\pi}{L_y}$ and $\psi_0=1$ for all simulations.
   The perturbed magnetic field is then
\beq
\mathbf{B'}= \frac{\psi_0}{k}[\sin{(kx)}\cos{(ky)}\mathbf{\hat x}-\cos{(kx)}\sin{(ky)}\mathbf{\hat y}].
\label{perturbb}
\eeq

Table ~\ref{modelparameters} summarizes selected runs.
 The A, I, C, and E prefixes indicate  adiabatic, isothermal,  uniform resistivity  ($\nu_e=0$),  and enhanced resistivity  ($\nu_e\ne 0$) respectively. 
Simulations IS and IL are equivalent to IE,  but with half  and twice  the domain size respectively.

The prefix W indicates simulations in which a dense plasma "wall" of zero temperature is injected at one end of the domain perpendicular to the $y$ direction  between $y=10.8$ and $y=12.8$  to facilitate a shock when the outflow is  super-magnetosonic. The wall is intended simply as a generic
obstacle. In a solar flare a natural obstacle arises in standard geometries as the downward directed component of the outflow from the X-point impinges upon the closed soft X-ray magnetic loop of previously reconnected field lines\cite{tsuneta}.

The wall is injected only once per simulation at time  $t_w$, 
 but we insert the wall  at different times for different runs to test the sensitivity to $t_w$.  The sound speed of  wall density is set to zero to prevent the wall from expanding due to pressure gradients.   
The injected wall must  supply a density contrast at least as high as that across any  shock it facilitates. The wall  density for the isothermal simulations was twice that of the adiabatic simulations, consistent with the fact that   the mass conservation jump condition 
demands a higher compression ratio across isothermal shocks. 
The five wall simulations with  enhanced resistivity $(\nu_e >0)$   led to stronger outflows than the two wall simulations with  $\nu_e=0$.

\section{Results}

\begin{figure}
\includegraphics[scale=.85]{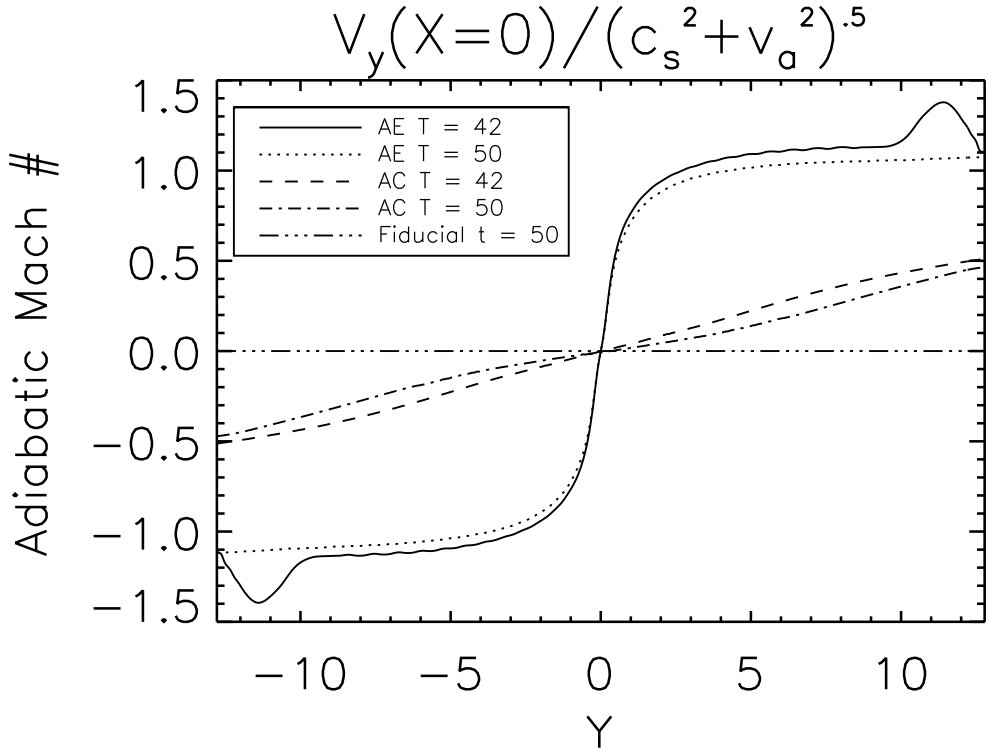}
\caption{Mach number $M=V_y/\sqrt{(c_s^2+v_A^2)}$  at $x=0$ for the adiabatic simulations AC and AE at times 42 and 50.  Constant resistivity simulation AC produced subsonic flows.  Simulation AE using enhanced resistivity produced moderately supersonic flows. The line marked 'fiducial' represents a  non-perturbed simulation which remained stable against reconnection and hence resulted in no flow velocty.}
\label{fig:adiaplot}
\end{figure}

\begin{figure}
\includegraphics[scale=.85]{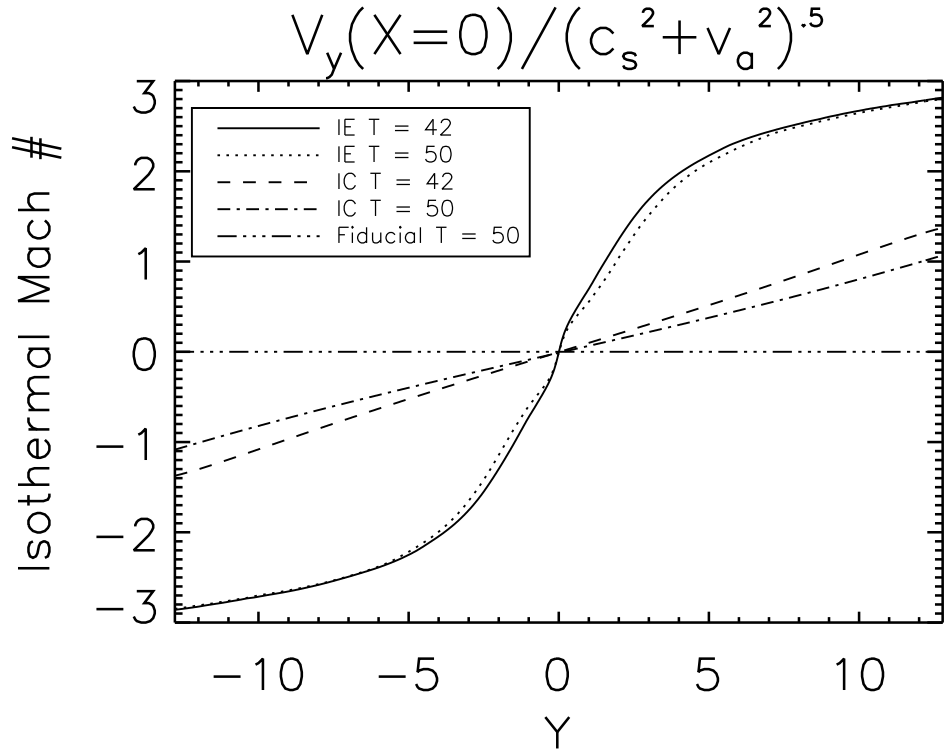}
\caption{Mach number $M=V_y/\sqrt{(c_s^2+v_A^2)}$  at $x=0$ for the isothermal simulations IC and IE at times 42 and 50.  Constant resistivity simulation IC produced marginally supersonic flows. Simulation IE using enhanced resistivity resulted ing reatly increased flow velocity.  One again, the non perturbed simulation marked as 'fiducial' showed stability against reconnection and no resulting flow.}
\label{fig:isoplot}
\end{figure}

Runs IC, IE, AC, \& AE
were  unstable to reconnection and resulted in accelerated outflows along the  x = 0 line.
For these wall-free runs, Table 1 shows that  at the final time $ t_f = 75$,
the uniform ($\nu_e =0$)  vs.  enhanced ($\nu_e/\nu_c =10$)  resistivity
cases  produce Mach numbers (given by $v_y/\sqrt{c_s^2+v_a^2}$) of  0.5 and 1.1 for the adiabatic simulations, 
 and 1.1 and 2.9 for the isothermal simulations. Figure (\ref{fig:adiaplot}) and Figure (\ref{fig:isoplot}) show the resulting flow profiles along the $x=0$ axis at times 42 and 50 and demonstrate that the flows had reached a steady state by $t=50$.  No flows were found to occur for test runs in which the perturbation to the magnetic field was not included.

The adiabatic simulations require $\nu_e >0$  to produce super-magnetosonic outflows but both adiabatic and isothermal cases produce higher outflow speeds with  $\nu_e >0$. 
These  trends are expected: First, isothermal flows imply a release of thermal energy from the system so that
the  conversion of magnetic to bulk ram pressure in the momentum equation is more efficient for isothermal flows, leading to a higher Mach number.  Heat is also released at isothermal shocks implying  higher compression ratios compared to the adiabatic cases. Second, when the central resistivity is enhanced, Petschek type "fast" reconnection
ensues \cite{biskamp00,kulsrud01}  albeit for the artificial incompressible limit, a centrally enhanced resistivity facilitates the rapid rotation of field lines through the plasma near the X-point required for the Petschek configuration\cite{kulsrud01}. The inflow  magnetic field exerts a tension force parallel to the outflow that helps accelerate the latter. Slow shocks (not to be confused with the outflow fast shocks)   appear in the inflow near the reconnection region for all our simulation cases where enhanced resistivity is used. This is shown in Figure (\ref{fig:shockimage}) and Figure (\ref{fig:shockimage2}).  Finally, lower mach number flows are expected in adiabatic simulations as the temperature of the plasma is increased which corresponds to an increase in the sound speed.

Although our present simulations are strictly MHD, since the ion-inertial length depends only on the plasma density, an enhanced resistivity would be expected for fully kinetic versions of our simulated 
parameter regime  based on a comparison between the Sweet-Parker thickness and the 
ion inertial length using such parameters:  For example, at $t = 50$ of a typical (e.g. isothermal) simulation 
we calculate the Sweet-Parker thickness ($\delta_{sp}= L/S^{1/2}$) where $S\equiv Lv_A/\nu_M$
Using  $L = 12.8$, 
inflow $v_A \sim 1.5$ (and $\beta \equiv c_s^2/v_A^2 \sim 0.2$)
and $\nu_c = 0.01$
we find 
$\delta_{sp}\sim 0.29<{c\over \Omega_{pi}}=1$ in our units. 
Plasmas with $\delta_{sp} < {c\over \omega_{pi}}$ (such as pre-flare solar  loops \cite{krucker07}),  
are  unstable to  enhanced dissipation near the X-point
\cite{biskamp00,kulsrud01,uzdensky03,uzdensky}.

\begin{figure}
\includegraphics[scale=.85]{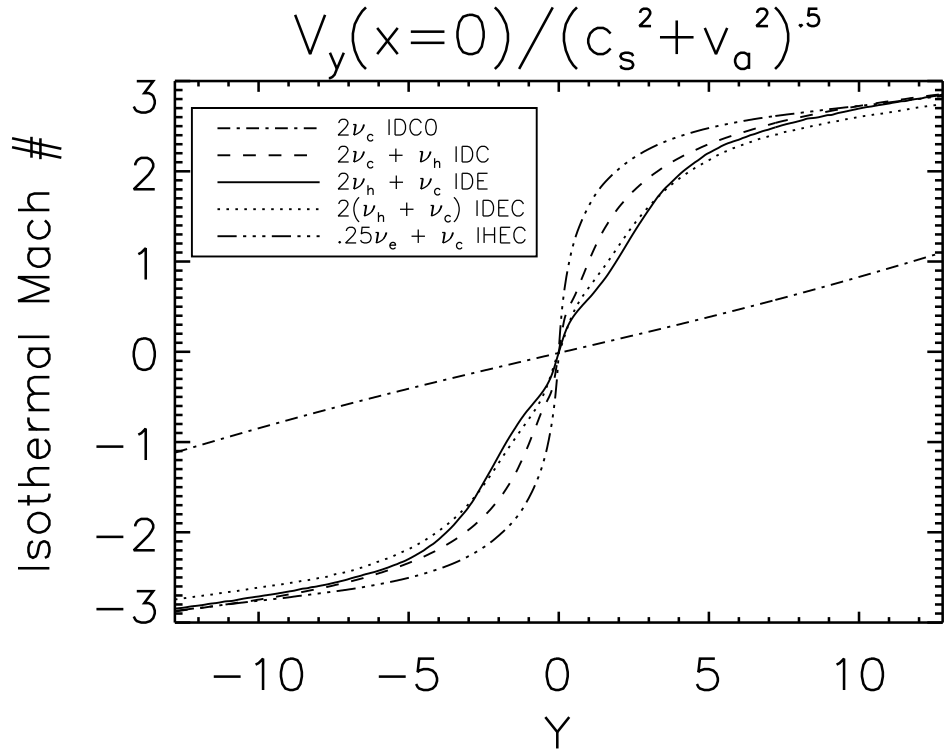}
\caption{Mach number $M=V_y/\sqrt{(c_s^2+v_A^2)}$  at $x=0$ for the isothermal simulations with varying resistivities.  The diagonal line represents uniform resistivity $\nu_c>0$
with $\nu_e=0$.  Doubling $\nu_c$ with $\nu_e=0$ leads to nearly the same line.
 In contrast, including enhanced resistivity $\nu_e>0$  with strengths ranging from 25\% to 200\% of $\nu_c$,  produce the curved profiles which become nearly  equivalent at large $y$.}
\label{fig:resist}
\end{figure}

Fig. \ref{fig:resist} shows the vertical flow profile at $t = 50$ for an isothermal EOS with varying combinations of $\nu_e$ and $\nu_c$ without the wall.  Isothermal simulations were used as the lack of heating allowed for a clearer study of the effects of resistivity on the flow profile.
The peak outflow  speed toward the edge of the box and the compression ratios are quite insensitive
 to    $\nu_e$  for $\nu_e >2.5\nu_c$, 
and insensitive to $\nu_c$ for $\nu_e=0$.

\begin{figure}
\includegraphics[scale=.85]{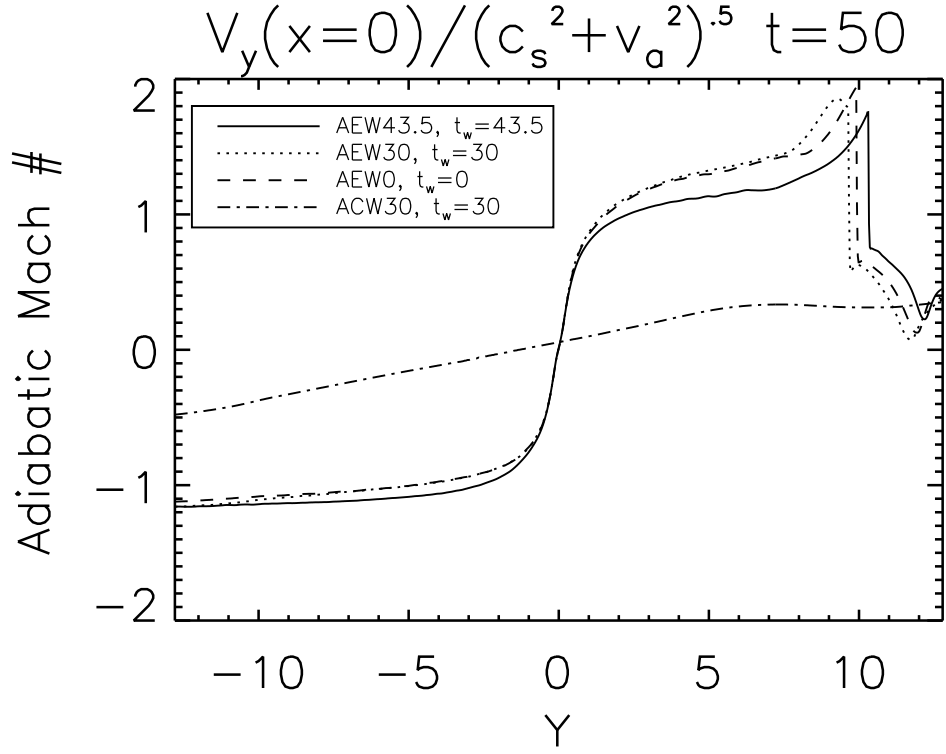}
\caption{Mach number $M=V_y/\sqrt{(c_s^2+v_A^2)}$ at $x=0$ and $t=50$ for the adiabatic simulations of Table 1 with walls placed near  $y\sim 11$ at times 0,  30, and 43.5. At negative $y$ there is no wall.  The curve that remains subsonic at all $y$ is run ACW30 in Table 1, the only case shown with $\nu_e=0$.}
\label{fig:shockplota50}
\end{figure}

Our main result is the formation of shocks in the  wall simulations (those with W in Table ~\ref{modelparameters}), across which we measured  compression ratios $r$  at $t=40, 45, 50$ along the y-axis.
Fig. \ref{fig:shockplota50} shows snapshots of the vertical flow along $x = 0$ at $t = 50$ for adiabatic $\nu_e> 0$ cases in which the wall was  inserted 
 at times, $t_w= 0, 30, 43.5$ (cases AEW0,  AEW30, \& AW43
 in Table ~\ref{modelparameters}) along with an  adiabatic $\nu_e=0$ case with wall inserted  at $t = 30$ (ACW30).  Only in cases with  $\nu_e/\nu_c > 2.5$,  did  a shock form, and in all such cases  maintained a compression ratio of $\sim 2$. The final strength of the shock is relatively insensitive to the time the wall was inserted and the shock location converged by $t = 50$ in all cases.  
 
\begin{figure}
\includegraphics[scale=.85 ]{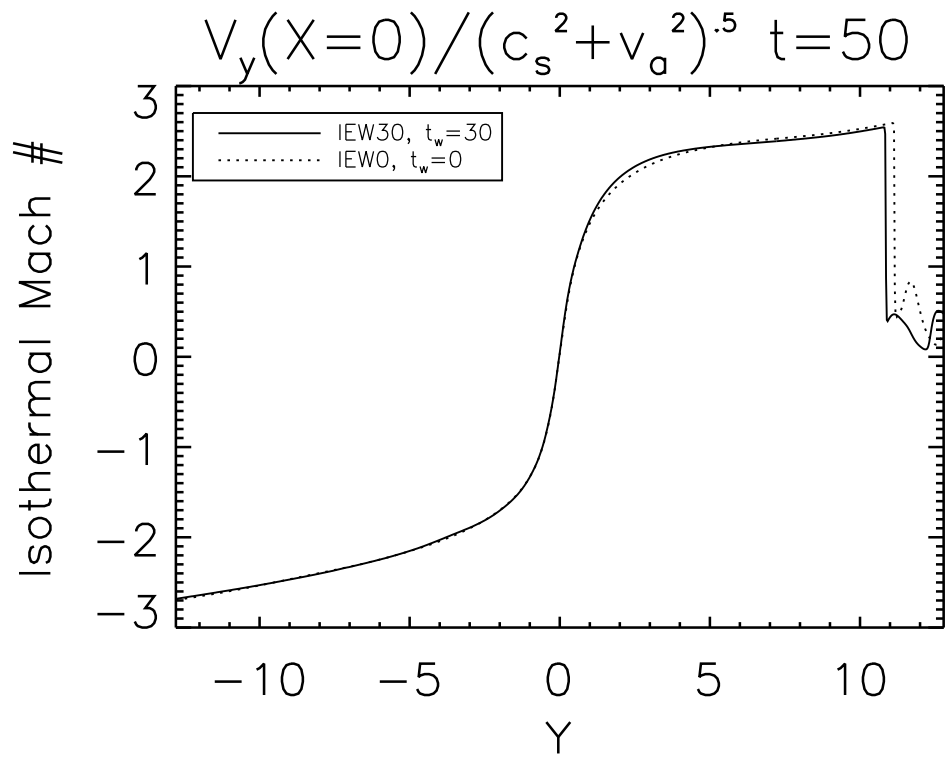}
\caption{Same as Fig. 2 for the isothermal simulations with walls  in Table 1.}
\label{fig:shockploti50}
\end{figure}

\begin{figure}
\includegraphics[scale=.40]{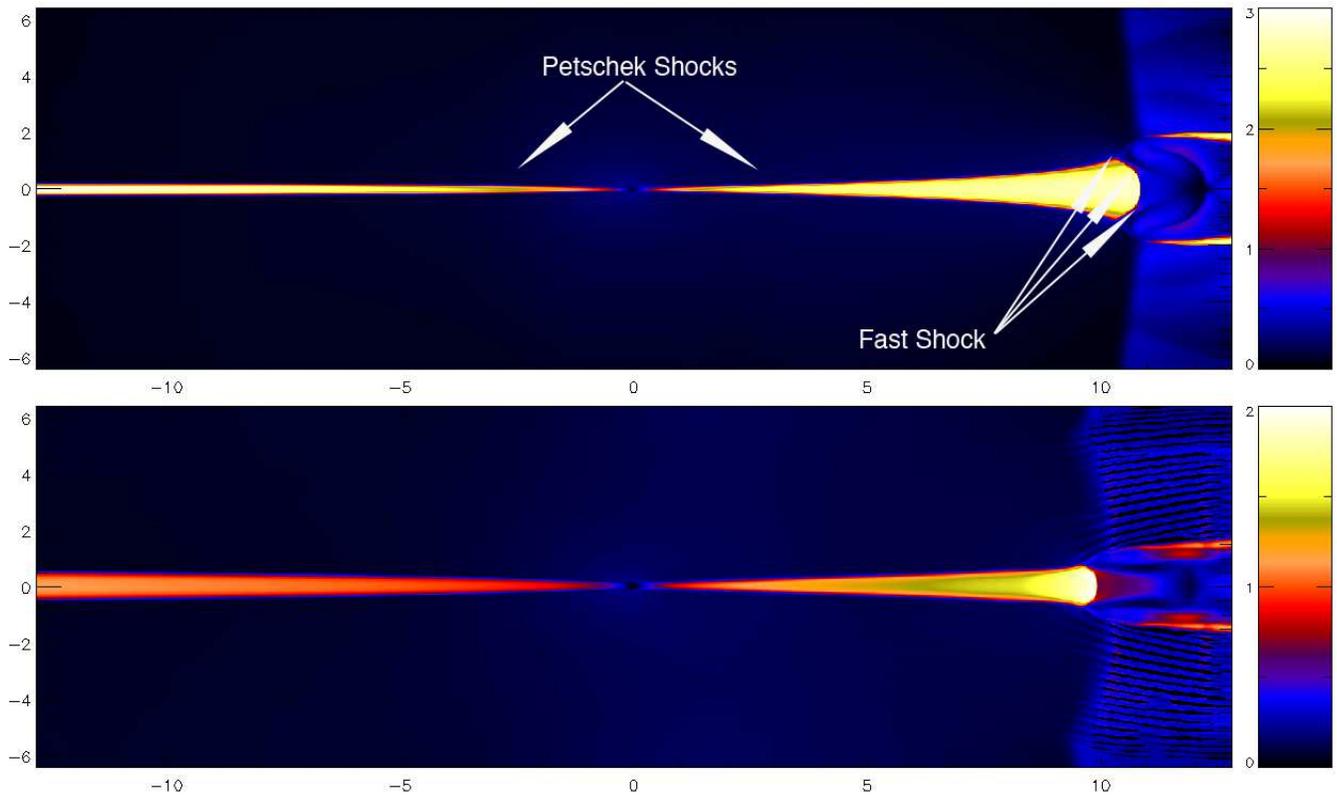}
\caption{Snapshot of the magenetosonic mach number for IEW30 (top image) and AEW30 (bottom image) at t = 50.   The color bars denote the magnitude of the mach number.  The location of the Petschek shocks and the fast shocks are labeled for the top figure only but occur at the same location for in the bottom figure.  (note the axes in this caption have been flipped to allow for easy plotting).} 
\label{fig:shockimage}
\end{figure}

\begin{figure}
\includegraphics[scale=.55]{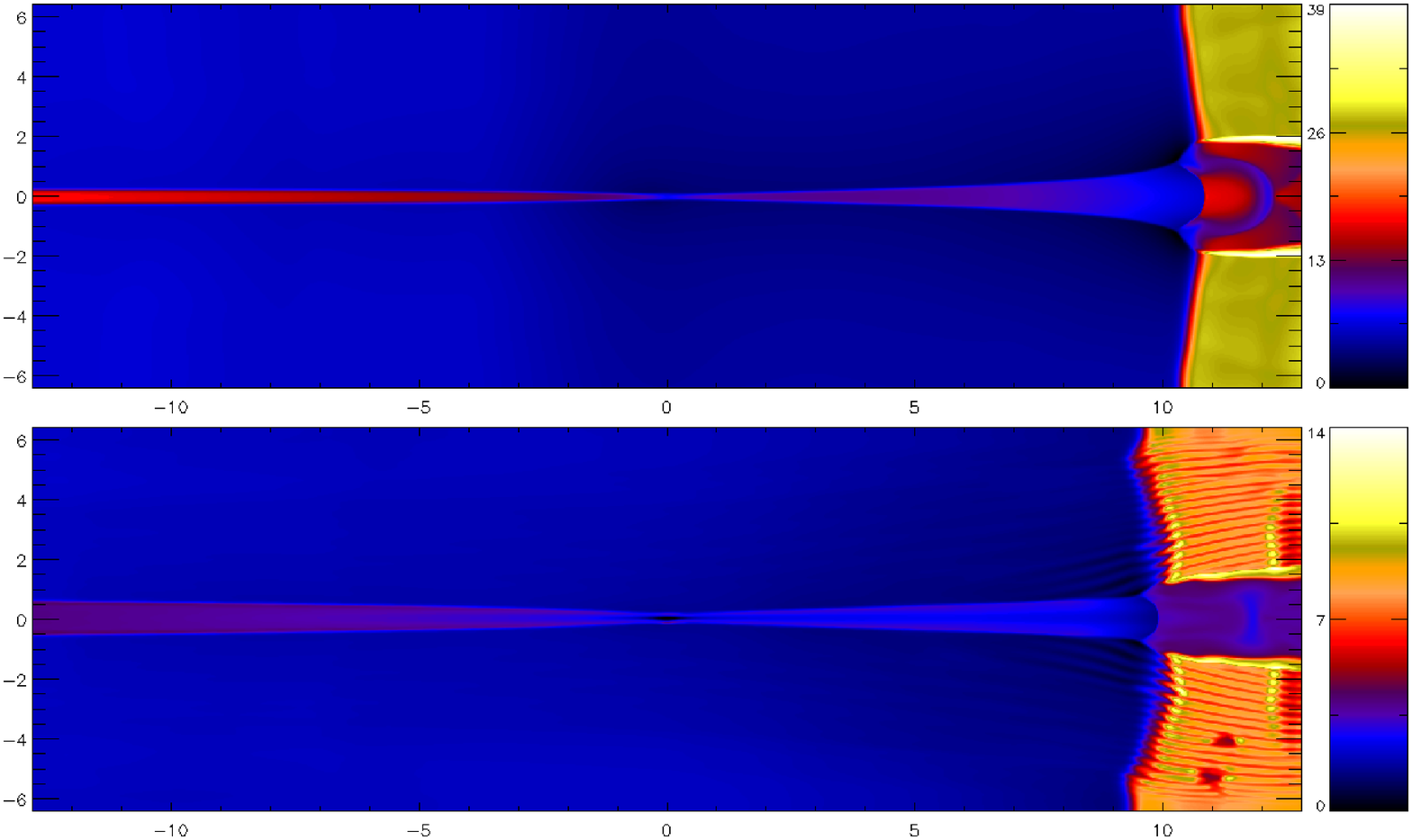}
\caption{Snapshot of the density profile for  IEW30 (top image) and AEW30 (bottom image) at t = 50.   The color bars denote the magnitude of the density.} 
\label{fig:shockimage2}
\end{figure}

Fig. \ref{fig:shockploti50} is analogous to  Fig. \ref{fig:shockplota50} for the analgous isothermal runs.   
 For $\nu_e>0$  with walls inserted at $t_w = 0$ and $30$, a shock formed with  compression ratio $\sim5.8\le r \le 6.8$ at $t = 45$ and $\sim 5.6$ at $t = 50$.

\section{Potential Implications for Shock-Fermi Acceleration and Flares}

In the limit that particle velocities are large compared to  shock velocities and both are non-relativistic, 
the analytically computed accelerated particle energy spectrum from shock-Fermi acceleration is \cite{bell2}
$N(E)\propto E^{-s}$, where $s={r+1/2\over r-1}$. 
For our adiabatic shocks  $r\sim 2$, so  $s \sim 2.5$.
(Note that a different proposed  first order Fermi acceleration  mechanism that also gives $s=2.5$
 was proposed in  \cite{dalpino}.  There, in the context of relativistic particle acceleration in active galactic nuclei, Fermi acceleration is argued to occur as the oncoming reconnection inflow streams flow toward each other.)

For a solar flare, 
the  index $s$ at emission sites is related to the  photon number spectral index  $\alpha$ via $s=\alpha + 2$,  (where $s-1$ is the electron number spectral index) based on emission from electron-ion interactions at magnetic loop foot-points\cite{the98}.
Electron energy spectral indices inferred from typical hard phases of flares
\cite{krucker07,gb08} are   $2\lesssim s \lesssim 3.5$, consistent with the adiabatic prediction of $s=2.5$ from our simulations 
 and consistent with scenarios  that invoke  fast shocks in reconnection down-flows\cite{blackman97,tsuneta98,blackman2}.

A typical flow travel time between the inferred X-point height and the loop top is $\sim 75$ seconds  \cite{tsuneta}, and the time scales for conductive losses \cite{tsuneta}
and bremsstrahlung cooling \cite{blackman2} for $\gtrsim 50$keV electrons are comparable or within an order of magnitude less. Any cooling that does incur would  make   $s$ inferred at the emission site  an upper limit to $s$ at the acceleration site.  Note that the shock itself would  be expected to be  adiabatic  because the shock thickness, determined by ion gyro-radii,  is much thinner than the cooling scale.

Having  chosen initial parameters to match  the GEM challenge\cite{birn01} for  consistency, we note that the distance bewteen the X-point and loop top obstacle in a typical flare   is  $>10^5$ times larger than our simulation box given the box density if we were to artificially scale dimensions into units of the ion-inertial length based on our chosen density.  However,  our  results
are strictly MHD and so the fact that the real solar flare is much larger in such dimensionless units
makes our conclusions that shocks are expected even  more robustly relevant for the solar flare scales since the existence of shocks does not depend on the density as an independent parameter once in the MHD regime.

In addition to the mere presence of shocks and the expected power law index, it is noteworthy that  the fast reconnection shocks in our simulations 
extend across  the entire outflow width with very little deviation in the compression ratio across along the shock (see Fig. 6).   All of the inflow then passes through the shock on its way out via mass conservation.   The efficiency of electron acceleration at the shock  would then not be  a question of geometric limitation but  a question of how efficiently the specific type of fast shocks  accelerate electrons. This contrasts, for example,  direct electric field electron acceleration along the x-line, which encounters  only a limited electron throughput.
A detailed study of the microphysics of shock acceleration at specifically weak fast shocks is an 
 ongoing PIC-simulation research topic that is  beyond the scope of the present work.
 The extent to which electrons must be "pre-injected" above a certain energy threshold \cite{JandE}
 to be accelerated at such shocks is not entirely clear.  More on this point below.
 
Observationally,  solar flares have different populations of electrons and different paradigms  have been proposed for the  potential role that  down stream fast shock acceleration
might play \cite{blackman1,blackman2,tsuneta97,tsuneta98}.  Some flares have more non-thermal electrons by fraction than others \cite{alexander97,emslie03}.  In compact-impulsive flares, which have a large percentage of non-thermal electrons, there is a knee  separating two power laws in the energy spectrum  \cite{dulk92} with the two power laws seemingly requiring different mechanisms \cite{blackman2}: For the particles within the higher energy steeper power law,  lower energy particles lag behind the higher energy particles, whereas, within  the lower energy power law electrons, the high energy particles lag the lower energy particles \cite{asch95,asch97}.  In the shock reprocessing paradigm \cite{blackman2}, the lower energy power law is taken to be the result of stochastic fermi acceleration in turbulent outflows from the reconnection X-point and 
and the highest energy particles are drawn from  those which subsequently get injected into the shock
as the flow passes through. The "obstacle" causing the shock
is a reconnected loop in the standard inverse-y picture of solar flares but 
the fraction of particles passing through the shock  in the outflow region then depends the convexity  of the loop; any pre-injection acceleration that might be needed for the particles to engage in the shock-Fermi acceleration could be accomplished by the stochastic acceleration region. 
In principle this can accommodate\cite{blackman2} the needed $\sim  10^{35}/s$ rate of electron throughput in typical flares \cite{masuda94}.

As we stated in the introduction, reconnection is best thought of as an acceleration environment
and the relative importance of   various mechanisms being possibly different in different flares.  
The discussion of the previous paragraph is one example of how such shocks may fit into
such a paradigm. However, the main point  of our paper here is  to highlight the robust 
appearance of such shocks in simulations and their generic presence,  regardless of whatever else may also be operating to produce observed phenomenology.

\section{Conclusions}

From 2-D MHD simulations, we have found  that super-magneotsonic outflows are a generic feature of fast magnetic reconnection simulations and produce shocks upon encountering a plasma wall. We used the same initial parameters as those of the GEM challenge MHD simulations\cite{birn01} to study this question.  We find the fast shocks in our adiabatic cases (the most relevant for solar flares) to be weakly compressive with $r < 2.5$.  Our results are generally consistent with previous analytic
predictions of the existence of weak fast mode shocks  in reconnection outflows 
\cite{blackman1} and consistent the use of such shocks for particle acceleration in  solar flares \cite{blackman97,tsuneta98,blackman2}.  The potential importance of such fast mode shocks
and the complexity of magnetic reconnection warranted our explicit demonstration of their formation in a reconnection numerical simulation.   It is also important to emphasize that the weak fast mode downstream shocks are distinct  from the Petschek slow mode upstream shocks.

If Fermi acceleration is operative at such shocks, it will  produce steeper particle energy spectra than  maximally compressive  $r=4$ adiabatic shocks. The resulting 
steeper particle spectral indices are not inconsistent with inferred distributions of solar flare electrons. An important  next step is to directly study the microphysics of particle acceleration at these weakly compressive fast mode reconnection outflow shocks;  most work has  focused on particle
acceleration at strong fast mode shocks.

We have not considered a guide field along the $z$-axis in the third dimension and depending on on the spatial gradient of of such a field the Mach numbers of the outflow may change. The influence of a guide field would likely depend on its initial spatial gradient as to whether its net effect is to enhance or reduce the outflow pressure gradient. For example,  if the guide field is very stronger at the X-point and weaker farther out, it could increase the outflow Mach number. If the field is uniform it would not produce a strong force. This is another topic for future work.

\begin{table}
\caption{Simulation runs. Column 2 is the resolution with symbols
 $I =240\times 480$, $II=384\times768$, $III=480
 \times 960$, $IV=960\times 1920$. The $t_f$ is the final simulation
 time, $M_p$ is the peak Mach number at the measured time, $r$ is the compression ratio
 across the shocks at the times indicated, and is relevant only for the  $M_p>1$ simulations
 in which a wall is injected at time $t_w$. NA indicates not applicable as no wall was present to generate a shock.
Magnetic diffusivity parameters $\nu_c,\nu_e$ and $\sigma$ are defined in the text. 
  }
\label{modelparameters}
\begin{tabular}{l | c | c | c | c | c | c|c|c|c|c} 
\hline\hline                        
Sim. \# & res. & 
 $\nu_c$  &$\nu_e/\nu_c$& $\sigma$ & $t_{w}$ & $t_{f}$&$M_p$  &$r(t=40/45/50)$\\ [0.5ex] 
\hline
AC  & $II$ &
 .01 & 0.0 & NA & NA & 75&0.46&NA\\
AE  & $II$ &  .01 & 10.0 &  0.25 & NA &75&1.07&NA\\
AEW0  & $III$ & .008 & 10.0 &  0.25 & 0.0 & 50&1.85&1.9/2.0/2.2\\
AEW30  & $III$ & .008 & 10.0 &  0.25 & 30.0 & 50&1.97&1.8/1.8/2.1\\
AEW43.5  & $III$ &  .008 & 10.0 &  0.25 & 43.5 & 50&1.79&NA/2.1/2.1\\
ACW30  & $III$ &  .008 & 0.0 &  0.25 & 30.0 & 50&0.35&0.0/0.0/0.0\\
IC & $II$ & .01 & 0.0 & NA & NA & 75&1.08&NA\\
IE & $II$ &  .01 & 10.0 &  0.25 & NA & 75&2.91&NA\\
IEW0  & $III$ &  .008 & 10.0 &  0.25 & 0.0 & 50&2.55&6.0/5.8/5.5\\
IEW30  & $III$ &  .008 & 10.0 &  0.25 & 30.0 & 50&2.6&7.0/6.8/5.7\\
IL&$IV$&  .008& 10&0.25&NA&50&2.86&NA\\
IS&$I$&  .008& 10&0.25&NA&50&2.85&NA\\
IDCO&$III$&  .016& 0&0&NA&50&1.1&NA\\
IDC&$III$&  .016& 10&0.25&NA&50&2.92&NA\\
IDE&$III$&  .008& 20&0.5&NA&50&2.92&NA\\
IDEC&$III$&  .016& 20&0.5&NA&50&2.83&NA\\
IHEC&$IV$&.008&2.5&.125&NA&50&2.83&NA\\

\hline
\end{tabular}
\end{table}

\begin{acknowledgments}
We thank J. Stone and K. Beckwith for invaluable support with the ATHENA code.
\end{acknowledgments}
\end{document}